\documentclass{aa}
\usepackage{graphicx}
\begin{document}
   \title{A photometric study of the two poorly known northern open
clusters NGC~133 and NGC~1348\thanks{Based on observations
carried out at Mt Ekar, Asiago, Italy},\thanks{Photometry is only available
in electronic form at the CDS via anonymous ftp to {\tt cdsarc.u-strasbg.fr (130.79.128.5} or via {\tt http://cdsweb.u-strasbg.fr/cgi-bin/qcat?J/A+A//}}}

   \author{Giovanni Carraro}

   \offprints{G. Carraro}

   \institute{
              Dipartimento di Astronomia, Universit\`a di Padova,
              Vicolo Osservatorio 2, I-35122 Padova, Italy\\
               \email{giovanni.carraro@unipd.it}
            }

   \date{Received February 2002; accepted}

   \abstract{ We present CCD UBVI observations obtained in the field
of the previously unstudied northern open clusters NGC~133 and
NGC~1348. We argue that NGC~133 is a heavily contaminated cluster, for
which we identify 13  candidate members down to $V$=14.50 mag 
on the basis of the position
in the two-color Diagram . Membership has been checked
against proper motions from Tycho~2, whenever available. The cluster
turns out to have a reddening E$(B-V)=0.60\pm0.10$ mag, to be 
$630\pm150$ pc distant from the Sun, and to have an age less than
10 Myrs. NGC~1348 is a more reddened clusters 
(E$(B-V)=0.85\pm0.15$ mag)  for which we isolate 20 members.
The cluster lies at a distance of $1.9\pm0.5$ kpc away from the Sun,
and has an age greater than 50 Myrs.

   }
\titlerunning{Photometry of NGC~133 and NGC~1348}

   \maketitle
%
%________________________________________________________________

\section{Introduction}
In this paper we present $UBVI$ photometry of two
northern open clusters, namely NGC~133 and NGC~1348,
for which we provide the first CCD study.\\
Both the clusters appear as weak concentrations
of a small group of bright stars well mixed with
the rich Galactic disk field star population toward
their direction.
This fact renders it difficult to study these objects,
and it is the main reason for which they have been
almost neglected up to now.\\
NGC~1348 has in fact never studied.
On the other hand, NGC~133 
(OCL~296, Lund~17, C0028+60, Trumpler class IV 1p:b)
seems to be  an asterism of 5 bright stars.
Evidence has been brought forth by 
an early photographic study by Jasevicius (1964, 1970)
that this  is not a cluster, but simply a random concentration
of a few bright stars.\\
In this study we would like to address the issue
of the real nature of these two clusters by means
of deep CCD multicolor photometry, and proper motion
data from Tycho~2 catalog.\\

This study is part of a long term project aimed
at providing accurate CCD photometry for
northern star clusters at the Asiago Astrophysical
Observatory (Carraro 2002, and references therein).\\

\noindent
The plan of this paper is as follows.\\
In Sect.~2 we briefly present the observations and data reduction.
Sects.~3 to 4 illustrate our results for
NGC~133 and NGC~1348, respectively.
Finally, Sect.~5 draws some conclusions and suggests further lines
of research.

\section{Observations and Data Reduction}

\begin{table}
\caption{Basic parameters of the observed objects.
Coordinates are for J2000.0 equinox}
\begin{tabular}{ccccc}
\hline
\hline
\multicolumn{1}{c}{Name} &
\multicolumn{1}{c}{$\alpha$}  &
\multicolumn{1}{c}{$\delta$}  &
\multicolumn{1}{c}{$l$} &
\multicolumn{1}{c}{$b$} \\
\hline
& {\rm $hh:mm:ss$} & {\rm $^{o}$~:~$^{\prime}$~:~$^{\prime\prime}$} & {\rm $^{o}$} & {\rm $^{o}$}\\
\hline
NGC~133        & 00:31:12 & +63:22:00 & 120.67 & +0.58\\
NGC~1348       & 03:34:10 & +51:24:46 & 146.97 & -3.70\\     
\hline\hline
\end{tabular}
\end{table}
Observations were carried out with the AFOSC camera at the 
1.82~m Copernico telescope of Cima Ekar (Asiago, Italy), in the photometric
night of December 18, 
2001. AFOSC samples a $8^\prime.14\times8^\prime.14$ field in a  
$1K\times 1K$ thinned CCD. The typical seeing was around 2.0  
arcsec.  
 
The basic data of the studied objects are summarized in Table~1, whereas
the details of the observations are listed in Table~2,
where the observed fields are reported together with the exposure
times, the typical seeing and the airmass.
The covered regions are shown in Figs.~1 and~8, where two XDSS\footnote
{Second generation Digital Sky Survey, {\tt http://cadcwww.dao.nrc.ca/cadcbin/getdss}} maps
are presented for NGC~133 and NGC~1348, 
respectively.
The data has been reduced by using the IRAF\footnote{IRAF  
is distributed by the National Optical Astronomy Observatories, 
which are operated by the Association of Universities for Research 
in Astronomy, Inc., under cooperative agreement with the National 
Science Foundation.} packages CCDRED, DAOPHOT, and PHOTCAL. 
The calibration equations obtained by observing Landolt(1992) 
SA~93, PG~1047+003, PG~2331+055 and PG~0231+051 fields along the night, are: 
	\begin{eqnarray}  
\nonumber 
u \! &=& \! U + 3.520\pm0.042 + (0.099\pm0.030)(U\!-\!B) + 0.58\,X \\  
\nonumber 
b \! &=& \! B + 1.407\pm0.012 - (0.004\pm0.017)(B\!-\!V) + 0.29\,X \\  
\nonumber 
v \! &=& \! V + 0.752\pm0.009 + (0.036\pm0.012)(B\!-\!V) + 0.16\,X \\  
\nonumber 
i \! &=& \! I + 1.619\pm0.017 - (0.011\pm0.015)(V\!-\!I) + 0.08\,X \\ 
	\label{eq_calib} 
	\end{eqnarray} 
where $UBVI$ are standard magnitudes, $ubvi$ are the instrumental  
ones, and $X$ is the airmass. The standard stars in these fields
provide a very good color coverage.
For the extinction coefficients, 
we assumed the typical values for the Asiago Observatory
(Desidera et al. 2001).
Photometric global errors have been estimated following Patat
\& Carraro (2001). For the $V$ filter,  
they amount at 0.03, 0.05 and 0.07 at $V\approx$
12.0, 16.0 and 20.0, respectively.

\begin{table} 
\tabcolsep 0.20truecm 
\caption{Journal of observations of NGC~133, NGC~1348,
and standard star fields (December~18, 2001).} 
\begin{tabular}{ccccc} 
\hline 
\multicolumn{1}{c}{Field}    & 
\multicolumn{1}{c}{Filter}    & 
\multicolumn{1}{c}{Time integration}& 
\multicolumn{1}{c}{Seeing}       &
\multicolumn{1}{c}{Airmass} \\
      &        & (sec)     & ($\prime\prime$)&\\ 
  
\hline 
 SA~93          &     &              &      &      \\
                & $U$ &  120         &  1.9 & 1.307\\
                & $B$ &  60,60       &  2.0 & 1.314\\ 
                & $V$ &  30,30,30    &  2.0 & 1.316\\ 
                & $I$ &  30,30       &  2.0 & 1.319\\ 
 NGC~133        &     &              &      &      \\ 
                & $U$ &  60          &  2.2 & 1.065\\ 
                & $B$ &  5,10,30     &  2.1 & 1.068\\ 
                & $V$ &  3,10.30     &  2.3 & 1.061\\ 
                & $I$ &  3,10        &  2.3 & 1.067\\
PG~1047+003     &     &              &      &      \\ 
                & $U$ &  120         &  1.9 & 1.245\\
                & $B$ &  60,60       &  2.0 & 1.240\\ 
                & $V$ &  30,30       &  2.0 & 1.250\\ 
                & $I$ &  30,30       &  2.0 & 1.238\\ 
PG~0231+051     &     &              &      &      \\ 
                & $U$ &  120         &  1.9 & 1.199\\
                & $B$ &  60,60       &  2.0 & 1.239\\ 
                & $V$ &  30,30       &  2.0 & 1.260\\ 
                & $I$ &  30,30       &  2.0 & 1.267\\ 
PG~2331+055     &     &              &      &      \\ 
                & $U$ &  120         &  1.9 & 1.109\\
                & $B$ &  60,60       &  2.0 & 1.108\\ 
                & $V$ &  30,30       &  2.0 & 1.107\\ 
                & $I$ &  30,30       &  2.1 & 1.107\\ 
 NGC~1348       &     &              &      &      \\ 
                & $U$ &  180         &  1.8 & 1.028\\ 
                & $B$ &  30,60       &  1.7 & 1.031\\ 
                & $V$ &  15,30       &  1.8 & 1.040\\ 
                & $I$ &  15,30       &  1.8 & 1.030\\
\hline 
\end{tabular} 
\end{table}

   \begin{figure}
   \centering
   \resizebox{\hsize}{!}{\includegraphics{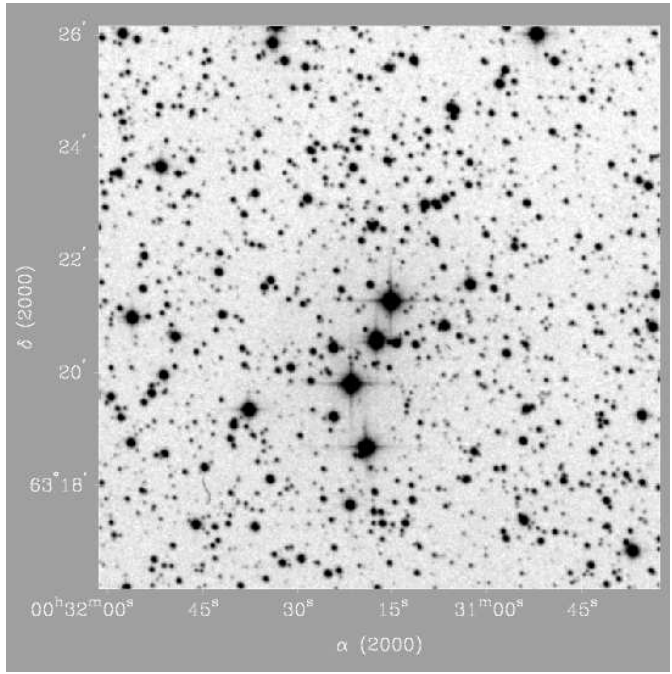}} 
   \caption{ A XDSS red map of the covered region in the field
of NGC~133. North is up, East on the left.}
    \end{figure}

\section{NGC~133}
NGC~133 
is identified by a group of 5 bright stars in a very
rich stellar field (see Fig.~1). 
One of these stars is the binary system BD+6293 (ADS~423A,B),
whose $A$ component is of $B3$ spectral type.
It is located very low in the
Galactic plane, and therefore presumably suffers from severe
field star contamination. This cluster was never studied before, but
for a photographic $UBV$ survey carried out
by Jasevicius (1964,1970). This study 
however does not report estimates of the cluster fundamental
parameters, but the author suggests that there is no cluster
in the direction of NGC~133.

   \begin{figure}
   \centering
   \resizebox{\hsize}{!}{\includegraphics{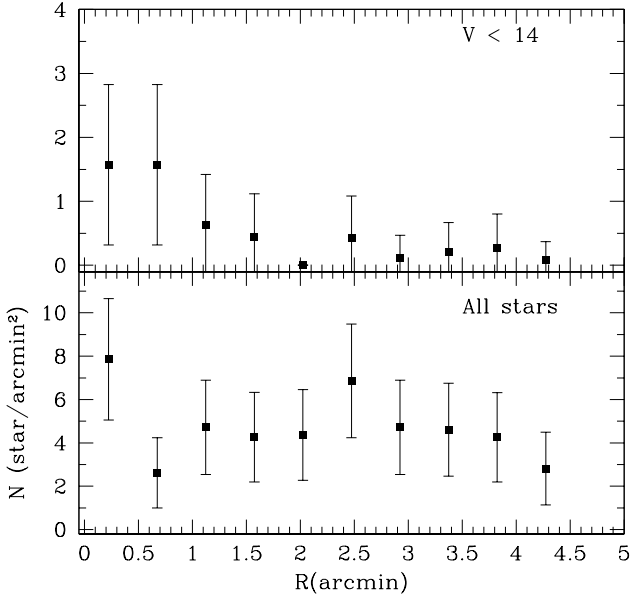}} 
   \caption{Star counts in the field of of NGC~133 as a function of
   the radius.{\bf Lower panel}: All the stars. {\bf Upper panel}:
   only stars brighter than V=14.} 
    \end{figure}

\subsection{Star counts}
According to Lyng\aa~ (1987), NGC~133 has a diameter of 7 arcmin,
so our study covers the entire cluster region.
To infer an estimate of the radius,
we derive  the surface stellar density by performing star counts
in concentric rings around the center of the covered area,
and then dividing by their
respective surfaces. The final density profile and the corresponding
Poisson error bars are depicted in Fig.~2.
In the lower panel  we take into account all the measured stars,
whereas in the upper panel we only consider stars brighter than
V = 14.
By inspecting the lower panel, one can readily see that
the profile decreases smoothly, and the cluster does not appear
as a clear concentration. This is probably due to the heavy 
contamination by faint Galactic field stars. In fact, after
selecting only the brightest stars, the cluster emerges more clearly
(see upper panel).
We estimate a cluster radius of about 2-2.5 arcmin, somewhat lower
than the value reported by Lyng\aa~(1987).

\begin{table*}
\tabcolsep 0.08cm
\caption{Photometry of likely member stars in the field of NGC~133
brighter than V=14}
\begin{tabular}{cccccccccccc}
\hline
\hline
\multicolumn{1}{c}{ID} &
\multicolumn{1}{c}{TYC 4019-} &
\multicolumn{1}{c}{Name} &
\multicolumn{1}{c}{$\alpha (J2000.0)$}&
\multicolumn{1}{c}{$\delta (J2000.0)$}&
\multicolumn{1}{c}{$V$}  &
\multicolumn{1}{c}{$(B-V)$} &
\multicolumn{1}{c}{$(U-B)$}  &
\multicolumn{1}{c}{$(V-I)$} &
\multicolumn{1}{c}{$\mu_\alpha cos \delta$ [mas/yr]} &
\multicolumn{1}{c}{$\mu_\delta$ [mas/yr]} &  
\multicolumn{1}{c}{E$(B-V)$}\\
\hline
   1&  2122&  HIP-2466& 00:31:14.9&+63:21:16.9&  9.501& 0.339&-0.083& 0.494&  0.0$\pm$2.8&-0.9$\pm$2.7&0.498\\
   4&  1038&  BD+6293 & 00:31:17.7&+63:20:33.0& 10.837& 0.404& 0.332& 0.456& -3.1$\pm$3.9&-2.8$\pm$3.9&0.499\\
   6&  1038&  BD+6293 & 00:31:17.5&+63:20:38.8& 11.572& 0.426& 0.405& 0.530&  2.2$\pm$3.7&-1.5$\pm$3.7&0.502\\
   7&  2326&          & 00:31:50.0&+63:23:54.1& 12.112& 0.323& 0.508& 1.527& -7.4$\pm$5.1&-1.2$\pm$5.3&0.540\\
   8&      &          & 00:31:02.2&+63:21:23.9& 12.444& 0.568& 0.493& 0.968&   &  &0.548\\
  10&      &          & 00:31:24.7&+63:30:27.4& 12.913& 0.542& 0.307& 0.695&   &  &0.579\\
  13&      &          & 00:31:07.2&+63:20:45.8& 13.076& 0.727& 0.547& 1.022&   &  &0.707\\
  15&      &          & 00:31:49.5&+63:20:51.9& 13.343& 0.485&-0.005& 0.782&   &  &0.614\\
  20&      &          & 00:31:17.6&+63:22:33.4& 14.076& 0.640& 0.034& 1.224&   &  &0.703\\
\hline
\end{tabular}
\end{table*}

\subsection{Proper motions}
Important information on the kinematics of the luminous stars in and around 
NGC~133 can be derived from the proper motions available in the Tycho-2 
catalogue. 
The Tycho-2 proper motions are based on the comparison between contemporary 
mean positions derived from the recent Tycho observations on-board Hipparcos 
and early-epoch positions observed many decades ago (see H{\o}g et al.\ 2000 
and references therein). 
Due to the long time-baseline they have rather high precision and 
therefore directly indicate the 
long-term mean tangential motions of the stars. 
We have collected proper motion components for 16 stars in a field
of $10^{\prime}.0 \times 10^{\prime}.0$ centered in NGC~133.
They are shown in the vector point diagram in
Fig.~3 together with the errors reported
in the Tycho~2 catalog. From this sample we have extracted 8 stars, 
which seem to crowd in the vector point diagram.
By assuming that these stars are likely members, we derive
the common mean motion (which we shall assume as the
cluster mean motion):

\[
\mu_\alpha = -0.7\pm4.7
\]

\[
\mu_\delta = -0.4\pm3.6  .
\]

\noindent
The errors reported in the Tycho~2 catalog amount at 
more than 2.5 mas/yr, and therefore we conclude that these 8 stars most
probably share a common tangential motion since their components
deviate less that 1$\sigma$ from the derived
mean motion. Therefore
in the direction of the object NGC~133 a star cluster
seems  to be present.

   \begin{figure}
   \centering
   \resizebox{\hsize}{!}{\includegraphics{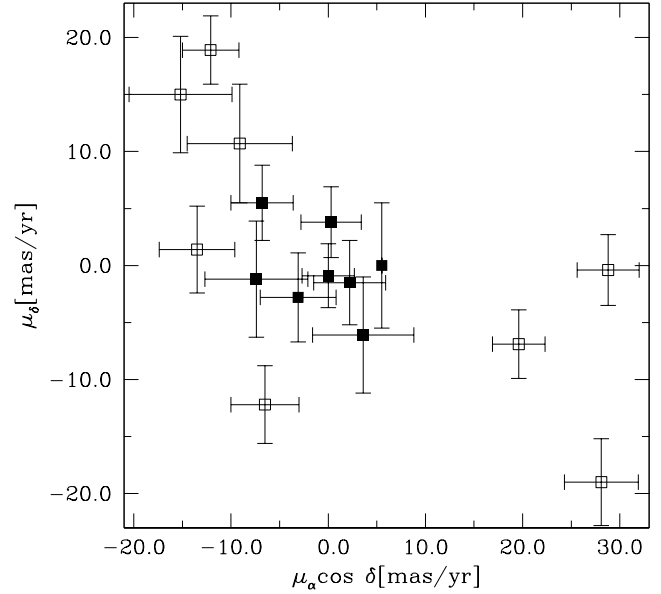}} 
   \caption{Vector point plot of Tycho-2 proper motion
and proper motion errors for the stars in the field of NGC~133.
Open symbols indicate likely non-members, filled symbols likely\
members.}
    \end{figure}

   \begin{figure*}
   \centering
   \resizebox{\hsize}{!}{\includegraphics{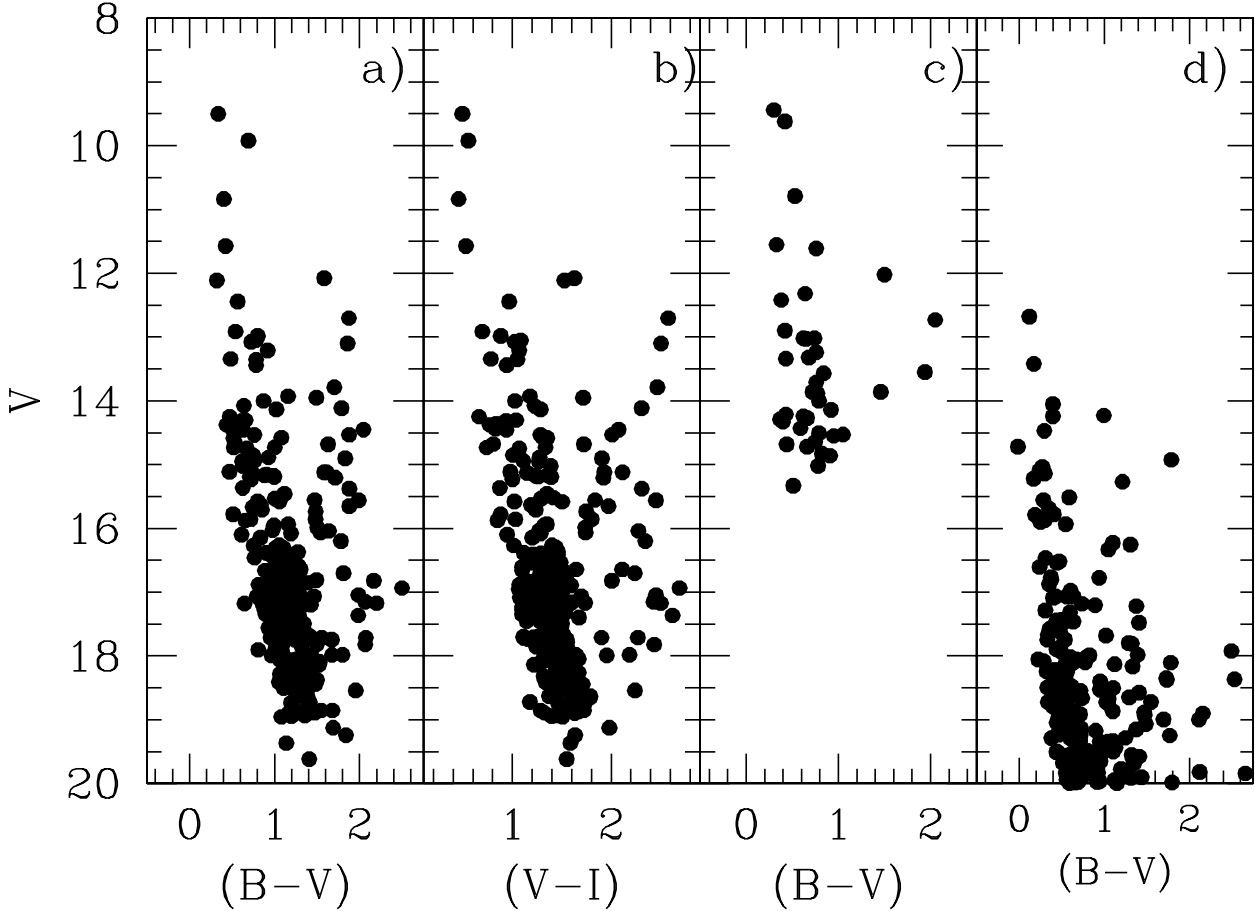}} 
   \caption{ CMDs of the stars in the region of NGC~133. {\bf
   panel a}:
    all the stars in the $V$ vs $(B-V)$ plane. {\bf Panel b}:
all the stars in the $V$ vs $(V-I)$ plane. {\bf Panel c}:
the CMD by Jasevicius (1964). {\bf Panel d}: A simulation
of the Galactic disk component in the direction of NGC~133.}
    \end{figure*}

\subsection{Color-Magnitude Diagrams}
The Color-Magnitude Diagrams (CMDs) for all the stars measured in the
direction of NGC~133 are shown in Fig.~4. In panel {\it a)}
we plot all the stars in the $V$ vs $(B-V)$ plane,
where in panel {\it b)} the same stars are plotted in the
$V$ vs $(V-I)$ plane. For comparison, the photographic
CMD by Jasevicius (1964) is presented panel in {\it c)}.
The comparison - in the sense (this study-Jasevicius) -
yields:

\[
\Delta V = 0.07\pm0.06 
\]

\[
\Delta (B-V) = 0.11\pm0.05
\]

\[
\Delta (U-B) = 0.13\pm0.09
\]

\noindent
for 29 common stars.\\
Clearly, the present study supersedes that of Jasevicius (1964).
The CMDs are not easy to interpret, since most of the stars
are just Galactic disk field stars. This is corroborated
by the CMD in panel {\it d)}, where a simulation is 
presented of the Galactic disk component toward NGC~133.
The simulation has been performed using the {\it TRILEGAL}
code (Girardi et al 2002), as calibrated by Groenewegen
et al (2002).
Also from this figure is evident that NGC~133 is a small
group of stars brighter than $V \approx 14$
above the mean stellar background.

   \begin{figure}
   \centering
   \resizebox{\hsize}{!}{\includegraphics{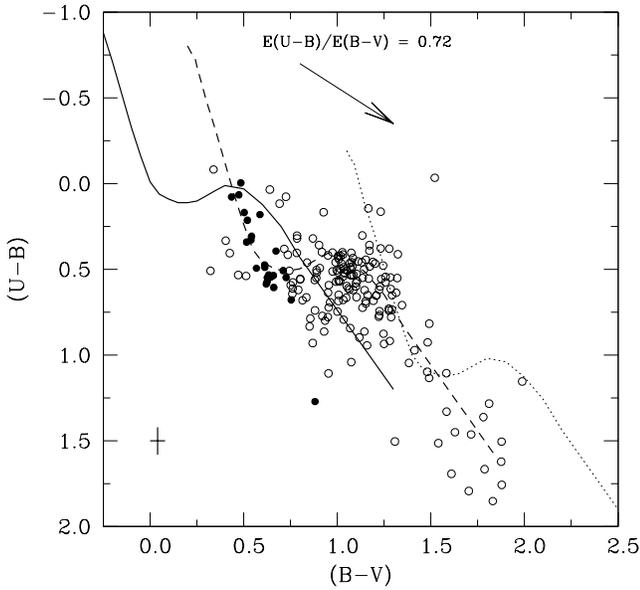}} 
   \caption{ Color-color diagram for all the stars in the field
of NGC~133 having $UBV$ photometry. 
The solid line is the  Schmidt-Kaler (1982) empirical ZAMS,
whereas the dashed and dotted lines are the same ZAMS, but shifted by 
E$(B-V)$~=~0.55 and 1.4, respectively. The cross indicates the typical error bars.}
    \end{figure}

   \begin{figure}
   \centering
   \resizebox{\hsize}{!}{\includegraphics{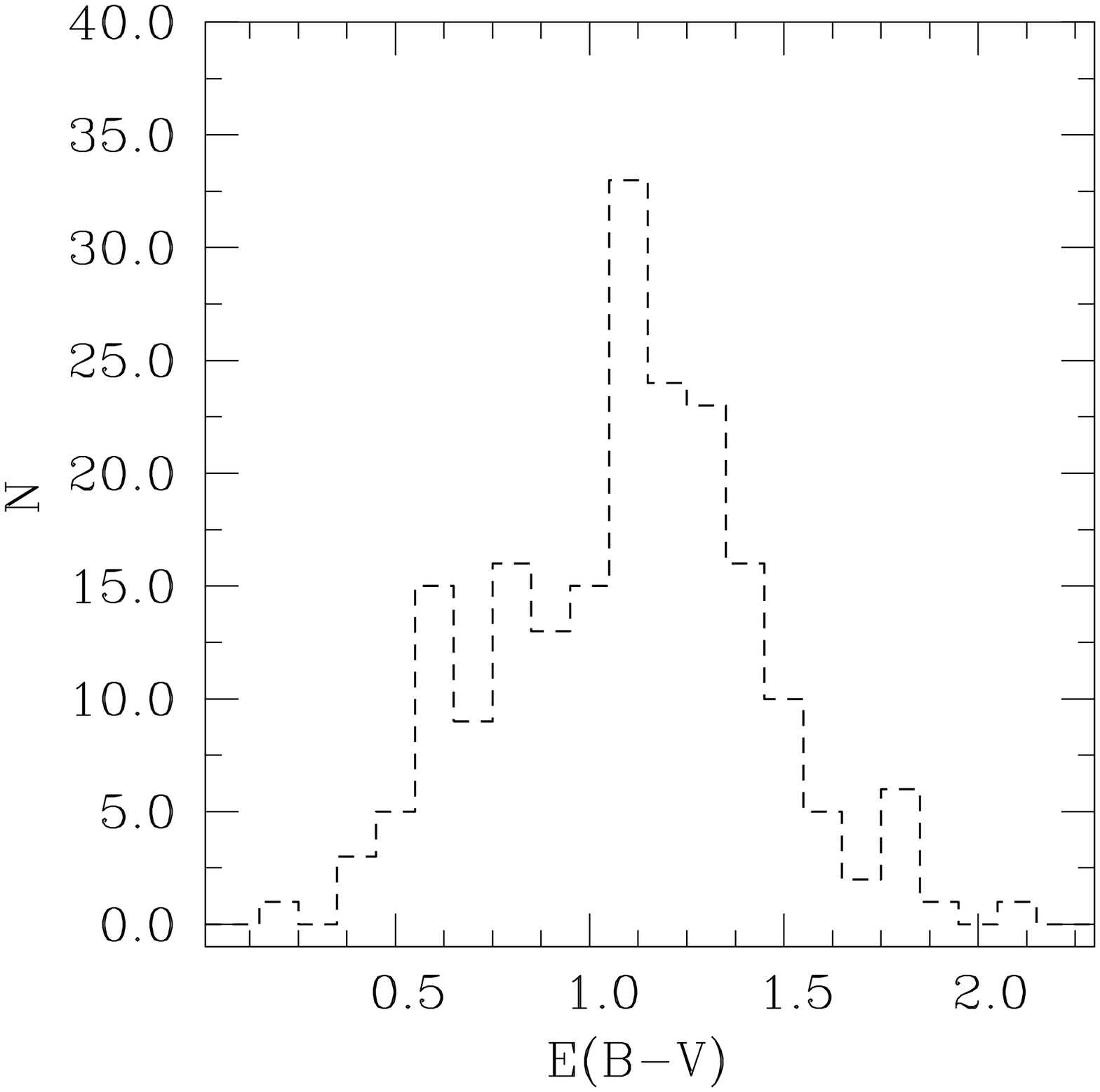}} 
   \caption{Reddening distribution for the stars in the region
of NGC~133 having $UBV$ photometry.} 
    \end{figure}

   \begin{figure}
   \centering
   \resizebox{\hsize}{!}{\includegraphics{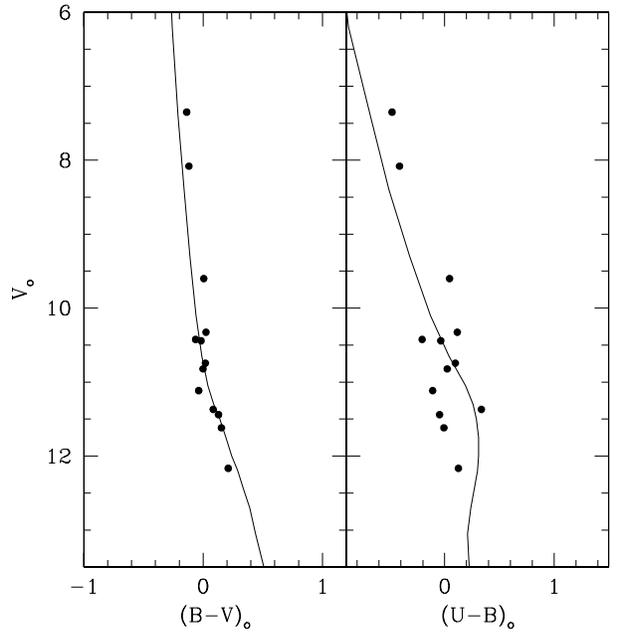}} 
   \caption{ Reddening corrected CMDs of the likely member
  stars in the region of NGC~133.}
    \end{figure}

\subsection{Two-color diagram and members selection}
We derive cluster membership by grouping stars according
to their mean reddening. Individual reddenings have been
computing by means of the usual reddening free parameter
$Q$, following the procedure outlined in detail
in Patat \& Carraro (2001).
The results are shown in Figs. 5 and 6.\\
In Fig.~5 we plot all the stars having $UBV$ photometry
in the two-color diagram. The solid line is an empirical
ZAMS from Schmidt-Kaler (1982).
There seem to be two pupulations. One having a mean
reddening E$(B-V)=0.60\pm0.10$ mag, which crowds
close to a ZAMS shifted by E$(B-V)=0.60$ mag (dashed line), and another
one with a much larger reddening. We consider
this latter population as the Galactic disk component,
made of stars placed at different distances, and
with a different amount of absorption. To guide the eye
we have drawn another ZAMS (dotted line) shifted by 
E$(B-V)=1.5$ mag. The same conclusion can be drawn by inspecting
Fig.~6, where we show the reddening distribution. 
This has a clear peak at E$(B-V) \approx 1.1$ mag, but at the same time
exhibits an hint for a  secondary peak at E$(B-V)=0.5-0.7$ mag.
However, the statistical significance of this secondary peak is
quite low, and has to be taken as no more than an indication.
An additional confirmation derives from Schlegel et al. (1998)
extinction maps, which in the direction of NGC~133
give $A_V = 1.426$ mag. By assuming a standard value
of the total to selective absorption ratio $R_V = 3.1$, the reddening
toward NGC~133 becomes E$(B-V) \approx 0.50$ mag,
in close agreement with our results.\\

\noindent
In conclusion, we would like to argue that the population
of stars having E$(B-V)=0.60\pm0.10$ mag (about 20 stars) 
identify the open cluster NGC~133.\\
Now, we need to compare these findings with the proper motion
data, to check for consistency.
Out of 16 stars which have Tycho~2 proper motions, 
we were able to secure photometry
only for 5 (see Table~3). In this Table, we list likely members
derived from the analysis of the two-color diagram brighter than $V \approx 14.0$.
The 4 stars with proper motion compatible with the
mean turn out to be also photometric members (see the last entry in
the table reporting the reddening), whereas
the star $\#2$  (Tycho 4019-744), which has
$\mu_{\alpha}=19.6$ mas/yr, $\mu_{\delta}=-6.9$ mas/yrs, turns out to
have E$(B-V)
=0.902$, which makes it both a photometric and an astrometric
non-member. This result makes us confindent when using photometrically
selected cluster members.\\
\noindent
It is however worth noting that these results have to be confirmed
by extending proper motion measurements to dimmer
magnitudes, and by providing radial velocities of the brightest
stars.

\subsection{Hints for NGC~133 distance and age}
In Fig.~7 we plot the reddening corrected CMDs for the likely
members stars. In both diagram we have over-imposed the empirical
Schmidt-Kaler (1982) ZAMS, shifted by $(m-M)_o =9.0\pm0.3$ mag,
which provides a nice fit of the stars distribution.
This implies that NGC~133 is located $630\pm150$ pc away from the Sun,
where the uncertainty mirrors the difficulty of the fit
due to the almost vertical structure of the MS.\\

\noindent
From the location of the stars in the $(B-V)$ vs $(U-B)$ plane,
we infer that the stars spectral types ranges from $B0$ to
$A5$ by deriving the absolute colors from the ZAMS at the same 
position of the stars. This result agrees with the $B3$
spectral type reported for ADS~423A ($\#4$ in the present numbering).
If the stars having $B0$ spectral type are still along the MS,
we derive an upper limit of 10 Myrs for the
age of NGC~133 (Girardi et al. 2000).

\section{NGC~1348}
This cluster was never studied before.
According to Lyng\aa~ (1987) it has a diameter of 5$^{\arcmin}$,
and the spectral type of the brightest stars is $A0$.
The region we covered is shown in Fig.~8. NGC~1348 is
identified by a weak over-density of stars in a rich stellar field.
Unfortunately, proper motions are not available for all
the stars in the region of NGC~1348, so we must rely only
on photometry to derive cluster members and cluster
fundamental parameters.

   \begin{figure}
   \centering
   \resizebox{\hsize}{!}{\includegraphics{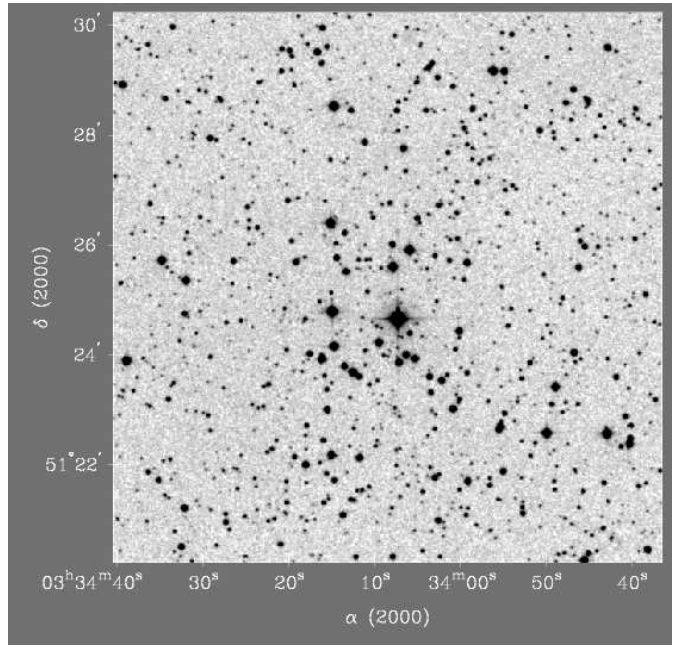}} 
   \caption{ A XDSS red map of the covered region in the field
of NGC~1348. North is up, East on the left.}
    \end{figure}

   \begin{figure}
   \centering
   \resizebox{\hsize}{!}{\includegraphics{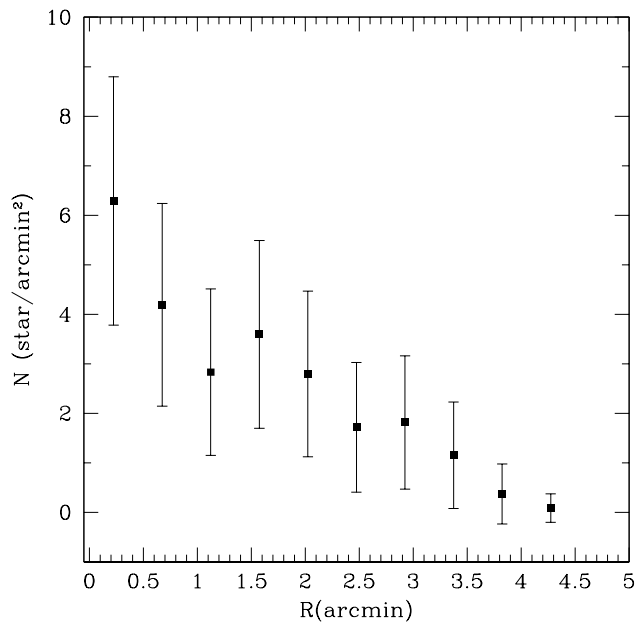}} 
   \caption{Star counts in the field of of NGC~1348 as a function of
   the radius.}
    \end{figure}

\subsection{Star counts}
We performed star counts in concentric rings around the brightest
star ($\#$1, Tycho~3325-48) in the field, assumed as the cluster center.
The number density profile is shown in Fig.~9. 
The profile decreases gently up to the limit of the covered region,
and therefore we suspect that the Lyng\aa~ estimate has to be taken
as a lower limit of NGC~1348 diameter, which would be at least
10$^{\prime}$.
Anyway, NGC~1348 appears as a real concentration.

   \begin{figure*}
   \centering
   \resizebox{\hsize}{!}{\includegraphics{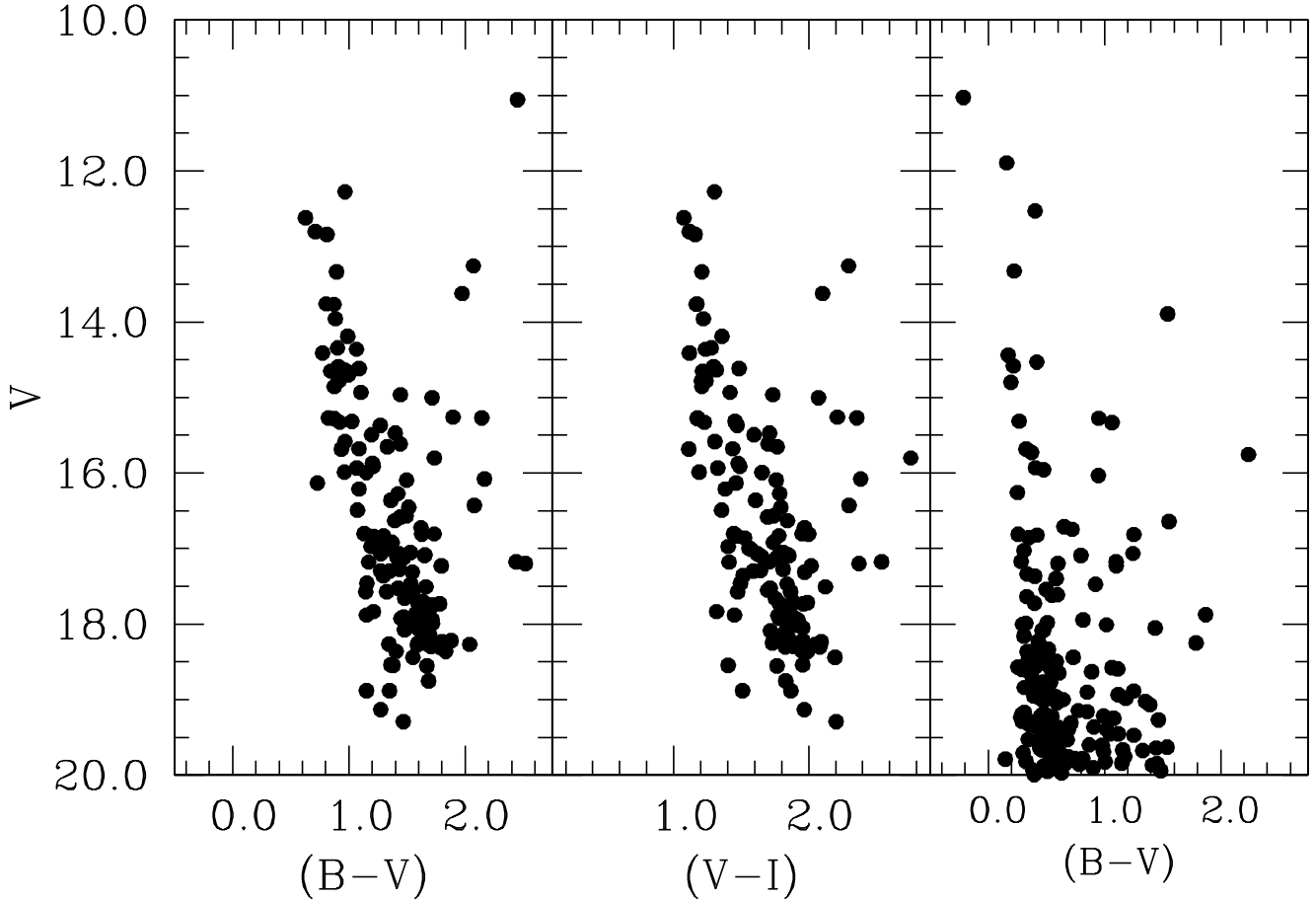}} 
   \caption{ CMDs of the stars in the region of NGC~1348. 
   {\bf Left panel}:
    all the stars in the $V$ vs $(B-V)$ plane. {\bf Central panel}:
all the stars in the $V$ vs $(V-I)$ plane.
{\bf Right panel}: A simulation
of the Galactic disk component in the direction of NGC~1348.}
    \end{figure*}

\subsection{Color-Magnitude Diagram} 
The CMDs for all the stars measured in the
direction of NGC~1348 is shown in Fig.~10. In the left panel 
we plot all the stars in the $V$ vs $(B-V)$ plane,
where in the middle  panel the same stars are plotted in the
$V$ vs $(V-I)$ plane. 
The CMDs are not easy to interpret, since most of the stars
are just Galactic disk field stars. This is confirmed
by the CMD in the right panel, where a simulation is 
presented of the Galactic disk component toward NGC~1348.
The simulation has been performed using the {\it TRILEGAL}
code (Girardi et al 2002), as calibrated by Groenewegen
et al (2002).
From this figure it is evident that NGC~1348 emerges as an 
overdensity of stars brighter 
than $V \approx 16-17$ above the mean stellar background.

\subsection{Two-color Diagram and member selection}
We follow the same method adopted above for NGC~133 to
derive individual reddenings and membership to the cluster.
The results are shown in Figs. 11 and 12.\\
In Fig.~11 we plot all the stars having $UBV$ photometry
in the two-color diagram. The solid line is an empirical
ZAMS from Schimdt-Kaler (1982).
There seems to be two populations. One having a mean
reddening E$(B-V)=0.85\pm0.15$ mag, which lies
close to a ZAMS shifted by E$(B-V)=0.85$ mag (dashed line), and another
one with much larger reddening. We consider
this latter population as the Galactic disk component,
made of stars located at different distances, and
with a different amount of absorption. To guide the eye
we have drawn another ZAMS (dotted line) shifted by 
E$(B-V)=1.5$ mag. The same conclusion can be drawn by inspecting
Fig.~12, where we show the reddening distribution. 
This has a clear peak at E$(B-V)=0.7-0.9$ mag, and several
smaller peaks at larger values of the reddening.\\

\noindent
We identify NGC~1348 with the group of stars having
reddening E$(B-V)=0.85\pm0.15$ mag (about 20 stars).

   \begin{figure}
   \centering
   \resizebox{\hsize}{!}{\includegraphics{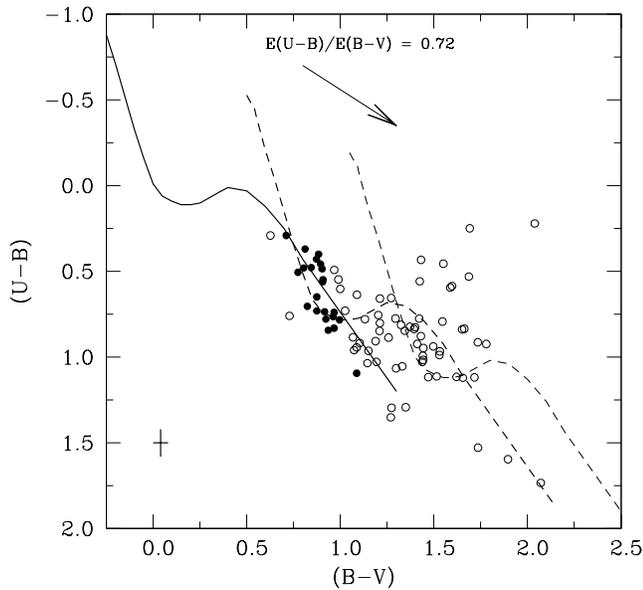}} 
   \caption{ Color-color diagram for all the stars in the field
of NGC~1348 having $UBV$ photometry. 
The solid line is the  Schmidt-Kaler (1982) empirical ZAMS,
whereas the dashed and dotted lines are the same ZAMS, but shifted by 
E$(B-V)$~=~0.85 and 1.5, respectively.The cross indicates the typical error bars.}
    \end{figure}

   \begin{figure}
   \centering
   \resizebox{\hsize}{!}{\includegraphics{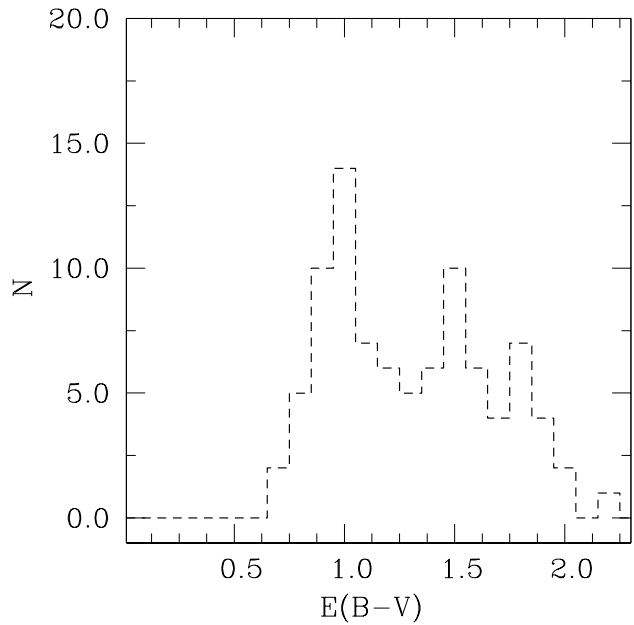}} 
   \caption{Reddening distribution for the stars in the region
of NGC~1348 having $UBV$ photometry.} 
    \end{figure}

\subsection{Hints for NGC~1348 distance and age}
In Fig.~13 we plot the reddening corrected CMDs for the likely
members stars above derived. 
In both diagrams we have over-imposed the empirical
Schmidt-Kaler (1982) ZAMS, shifted by $(m-M)_o = 11.5\pm0.5$ mag,
which provides a nice fit to the stellar distribution.
This implies that NGC~1348 is located $1.9\pm0.4$ kpc away from the Sun,
where the uncertainty mirrors the difficulty of the fit
due to the almost vertical structure of the MS.\\

\noindent
From the location of the stars in the $(B-V)$ vs $(U-B)$ plane,
we infer that the stars spectral types ranges from $B5$ to
$A5$ by deriving the absolute colors from the ZAMS at the same 
position of the stars.
Moreover by inspecting Fig~13, one can readily see that the brighter
stars are actually leaving the MS, whereas
the stars at $V_o \approx 11.00$ - with spectral types in the range
$B8-A0$ - are much
probably still on the MS. Therefore  
we derive a lower limit of 50 Myrs for the age of NGC~1348
(Girardi et al. 2000).

   \begin{figure}
   \centering
   \resizebox{\hsize}{!}{\includegraphics{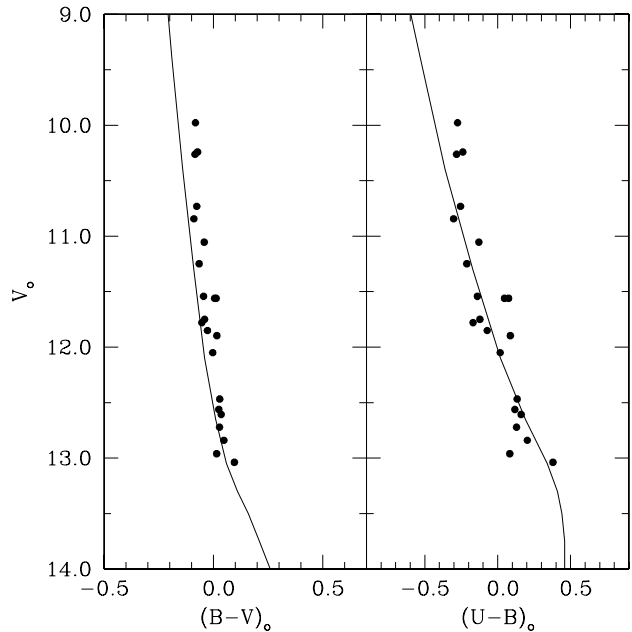}} 
   \caption{ Reddening corrected CMDs of the likely member
  stars in the region of NGC~1348.}
    \end{figure}

\section{Conclusions}
We have presented the first multicolor CCD study of 
the two poorly known northern open clusters NGC~133 and NGC~1348.
In the case of NGC~1348, no investigation had been carried
out insofar.\\
Our analysis shows that these objects are two poorly
populated, highly contaminated
clusters. In details, we find that:

\begin{itemize}
\item NGC~133 is a 10-15 stars group located about 600 pc
from the Sun. We estimate a reddening $E(B-V)=0.60\pm0.10$
and a probable age of less than 10 Myrs;
\item NGC~1348 is a 20 stars group with a mean reddening
$E(B-V)=0.85\pm0.15$, a distance of almost 2 kpc, and
an age greater than 50 Myrs.
\end{itemize}

\noindent
The results of this study are hampered by the strong field
star contamination.
Much better constrains on the nature and basic parameters
of these clusters can be obtained by enlarging the number of
stars with proper motions measurements, and by
obtaining radial velocity and spectral classification
of the brightest stars.

\begin{acknowledgements}
I deeply thank dr. Saulius Raudeliunas for providing
me with a copy of V. Jasevicius papers.
The kind night assistance at Asiago Observatory by Gigi
Lessio and Silvano Desidera is warmly acknowledged.
It is a real pleasure to thank Leo Girardi for his
permission to use the {\it TRILEGAL} code in advance
of publication.
This study made use of Simbad and WEBDA catalogs. 
\end{acknowledgements}

\end{document}